\documentclass[twocolumn,superscriptaddress,showpacs,preprintnumbers,amsmath,amssymb,prl]{revtex4}

\usepackage{graphicx}
\usepackage{dcolumn}
\usepackage{bm}

\begin{document}

\title{Anomalous enhancement of quasiparticle current near a potential
barrier in a Bose-Einstein condensate}

\author{Shunji Tsuchiya}
\author{Yoji Ohashi}%
\affiliation{Department of Physics, Keio University, 3-14-1 Hiyoshi, Kohoku-ku, Yokohama 223-8522, Japan}
\affiliation{CREST(JST), 4-1-8 Honcho, Saitama 332-0012, Japan}

\date{\today}

\begin{abstract}
We investigate tunneling properties of Bogoliubov phonons in a
Bose-Einstein condensate. 
We find the anomalous enhancement of the quasiparticle current $J_{\rm q}$ carried by
Bogoliubov phonons near a potential barrier, due to the
supply of the excess current from the condensate. This effect leads to
the increase of quasiparticle transmission probability in the low
energy region found by Kovrizhin {\it et al.}.
We also show that the quasiparticle current twists the phase of the
condensate wavefunction across the barrier, leading to a finite Josephson
supercurrent $J_{\rm s}$ through the barrier. This induced supercurrent
flows in the opposite direction to the quasiparticle current so as to
cancel out the enhancement of $J_{\rm q}$ and conserve the total current
 $J=J_{\rm q}+J_{\rm s}$.

\end{abstract}

\maketitle

Phonon in a superfluid system is an example of Goldstone mode which
appears in various fields of physics associated with spontaneous broken
symmetry \cite{Anderson}.
It is a manifestation of broken (global) $U(1)$ symmetry which underlies the
macroscopic quantum nature of the system, and a key to understand the low
energy properties of superfluids. In particular, in a Bose superfluid, it
plays fundamental roles for the superfluidity \cite{Landau}.
Phonon has been observed in various systems such as superfluid $^4$He \cite{Donnelly},
superconducting films \cite{Carlson},
atomic superfluid fermi gases \cite{Joseph},
as well as Bose-Einstein condensation (BEC) of cold atomic gases
\cite{Andrews}.
Due to the high degree of controllability, BEC of cold
atomic gases offers an opportunity to study novel properties of phonons in
the superfluid phase.

Bogoliubov phonon \cite{Bogoliubov} in a BEC has been a long-standing
issue of investigation
in cold atomic gases \cite{Davidson}.
In the last few years, quantum tunneling of Bogoliubov phonon has
attracted much attention \cite{Kovrizhin, Kagan, Danshita1, Danshita2,
Kato, Padmore}.
Since Bogoliubov phonon is a collective excitation of a BEC,
its tunneling property has specific features which are quite different from that of free particles.
In fact, an anomalous tunneling property of Bogoliubov phonon has been
predicted in \cite{Kovrizhin,Kagan}. 
It has been shown that the transmission probability of Bogoliubov phonon
through a potential barrier increases at low energies and always unity
in the zero-energy limit, irrespective of the height of the barrier
\cite{Kovrizhin, Kagan}.
Although several mechanisms were proposed in \cite{Kagan,Kato}, the
underlying physics of this anomalous tunneling has not been understood yet.

In this paper, we report alternative anomalous tunneling properties of
Bogoliubov phonons.
We find that the quasiparticle current is
{\it not} conserved, but greatly enhanced near the potential
barrier at low incident energies, due to the excess current supplied from the condensate.
This anomalous enhancement of the quasiparticle current increases the transmission
probability of Bogoliubov phonon in the low energy region, which
is consistent with the tunneling property in \cite{Kovrizhin, Kagan}.
In addition, we show that the quasiparticle current twists the phase of the
condensate wavefunction, leading to a Josephson supercurrent through the barrier. 
The excess part of the quasiparticle current is canceled out by this
induced supercurrent, so that the total current, given by the sum of the
quasiparticle current and supercurrent, is conserved.

We consider a tunneling of a Bogoliubov phonon at $T=0$ through a potential
barrier which only depends on $x$. 
This one-dimensional potential barrier was used in a recent experiment \cite{Engels}.
Ignoring the motion of atoms in the
$y$- and $z$-direction, we can treat this model as a one-dimensional
problem. For simplicity, we ignore effects of a harmonic trap.
This is allowed in a box-shaped trap \cite{Meyrath}.
To describe the Bose-condensed phase, we divide the boson field
operator $\hat\psi(x)$ into the sum of the BEC order parameter
$\Psi_0(x)$ and the non-condensate part $\delta\hat\psi(x)$ \cite{Pitaevskii}, as
$\hat\psi(x)=\Psi_0(x)+\delta\hat\psi(x)$.
The condensate wavefunction $\Psi_0(x)=\langle \hat\psi(x)\rangle$ obeys
the static Gross-Pitaevskii (GP) equation, given by (setting $\hbar=1$)
\begin{eqnarray}
\left(-\frac{1}{2m}\frac{d^2}{dx^2}+U(x)+g|\Psi_0|^2\right)\Psi_0=\mu
 \Psi_0\ ,
\label{GP}
\end{eqnarray}
where $m$, $\mu$, and $U(x)$ represent the mass of a boson, chemical
potential, and a potential barrier, respectively. $g\equiv 4\pi a/m$ is
the interaction between bosons, where $a(>0)$ is the $s$-wave scattering
length.

In the Bogoliubov mean-field theory, the non-condensate part has the form,
$\delta\hat\psi=\sum_j[u_j(x)\hat\alpha_j-v_j(x)^\ast\hat\alpha_j^\dagger]$.
Here, $\hat\alpha_j^\dagger$ ($\hat\alpha_j$) is the creation
(annihilation) operator of an excitation in the $j$-th state,
satisfying the bosonic commutation relation
$[\hat\alpha_i,\hat\alpha_j^\dagger]=\delta_{ij}$.
$u_j$ and $v_j$ are determined by the following Bogoliubov equation:
\begin{eqnarray}
\left(
\begin{matrix}
\hat{h} & -g\Psi_0^2\\
-g(\Psi_0^\ast)^2 & \hat{h}
\end{matrix}
\right)
\left(
\begin{array}{l}
u_j\\
v_j
\end{array}
\right)
=
E_j\tau_3
\left(
\begin{array}{l}
u_j\\
v_j
\end{array}
\right),\label{Bogoliubov}
\end{eqnarray}
where $\hat h \equiv
-\frac{1}{2m}\frac{d^2}{dx^2}+U(x)+2g|\Psi_0|^2-\mu$. $\tau_3$ is the
Pauli matrix, and $E_j$ is the Bogoliubov excitation spectrum. 

We consider a simple rectangular potential barrier
$U(x)=U_0\theta(d/2-|x|)$ ($U_0>0$). 
In the absence of supercurrent, we can safely take the condensate
wavefunction $\Psi_0$ to be real. In this case, the analytic solution of
Eq.~(\ref{GP}) is given in Ref.~\cite{Kagan}, as 
$\Psi_0(x) = \sqrt{n_0}\tanh\left[(|x|-d/2)/\sqrt{2}\xi+{\rm
arctanh}\gamma\right]$ ($|x|>d/2$), and 
$\Psi_0(x)=\sqrt{n_0}\beta/{\rm cn}(\sqrt{K^2+\beta^2}x/\sqrt{2}\xi,q)$
($|x|<d/2$), where ${\rm cn}(x,q)$ is the Jacobi elliptic function,
$\beta\equiv \Psi_0(0)/\sqrt{n_0}$,
$\gamma\equiv\Psi_0(d/2)/\sqrt{n_0}$, 
$K\equiv\sqrt{\beta^2+2(U_0/\mu-1)}$, and $q\equiv
K/\sqrt{K^2+\beta^2}$. 
$n_0\equiv \mu/g$ is the condensate density far away from the barrier
($x\to\pm\infty$), and $\xi\equiv 1/\sqrt{2mgn_0}$ is the healing length. 

In the absence of the barrier, Eqs. (\ref{GP}) and (\ref{Bogoliubov})
give the chemical potential $\mu=gn_0$ and excitation energy
$E=\sqrt{\varepsilon_p(\varepsilon_p+2gn_0)}$ (where
$\varepsilon_p=p^2/2m$). The wavefunction has the form
$(u(x),v(x))=(u_p,v_p)e^{ipx}$, where $u_p$ and $v_p$ are given by
\begin{eqnarray}
\left(
\begin{array}{l}
u_p\\
v_p
\end{array}
\right)=
\left(
\begin{array}{l}
a\\
b
\end{array}
\right)=
\left(
\begin{array}{l}
\frac{1}{\sqrt{2V}}\sqrt{\frac{\varepsilon_p+gn_0}{E}+1}\\
\frac{1}{\sqrt{2V}}\sqrt{\frac{\varepsilon_p+gn_0}{E}-1}
\end{array}
\right),
\label{eq.2}
\end{eqnarray}
where $V$ is the volume of the system.
$p=\pm k\equiv\pm\sqrt{2m}\sqrt{\sqrt{E^2+(gn_0)^2}-gn_0}$ describe 
the ordinary propagating waves in the $\pm x$-direction. 
In considering an inhomogeneous system, however, we note that, besides
the propagating solutions in Eq.~(\ref{eq.2}), Eq.~(\ref{Bogoliubov})
also has other two localized solutions, having the form of
$(u_p,v_p)=(-b,a)$, where $p=\pm i\kappa\equiv\pm
i\sqrt{2m}\sqrt{\sqrt{E^2+(gn_0)^2}+gn_0}$. 
The normalization of these localized states is given by $u_p^2-v_p^2=-1/V$. 
When the incident Bogoliubov phonon comes from $x=-\infty$, the
asymptotic solution is given by
\begin{eqnarray}
\left\{
\begin{array}{ll}
\left(\begin{array}{l}
u\\
v
\end{array}\right)=
\left(\begin{array}{l}
a\\
b
\end{array}\right)e^{ikx}+
r\left(
\begin{array}{l}
a\\
b
\end{array}\right)e^{-ikx}+
A\left(\begin{array}{l}
-b\\
a
\end{array}
\right)e^{\kappa x},\\
\ \qquad\qquad\qquad\qquad\qquad\qquad\qquad\qquad(x\to-\infty)\ ,
\\
\left(\begin{array}{l}
u\\
v
\end{array}\right)=
t\left(
\begin{array}{l}
a\\
b
\end{array}
\right)e^{ikx}+
B\left(
\begin{array}{l}
-b\\
a\end{array}
\right)e^{-\kappa x},\\
\ \qquad\qquad\qquad\qquad\qquad\qquad\qquad\qquad(x\to\infty)\ .
\end{array}
\right.
\label{asympt}
\end{eqnarray}
Here, $r$ and $t$ are, respectively, the reflection and transmission
amplitudes, satisfying $|r|^2+|t|^2=1$. As discussed later, this
condition results from the conservation of energy flux. 
In Eq.~(\ref{asympt}), $A$ and $B$ are the amplitudes of the localized
components near the potential barrier. 

We numerically solve the Bogoliubov equation Eq.~(\ref{Bogoliubov}) so
as to satisfy the asymptotic solution in Eq.~(\ref{asympt}). 
For this purpose, we employ the finite element method. 

\begin{figure}
\centerline{\includegraphics[width=6cm]{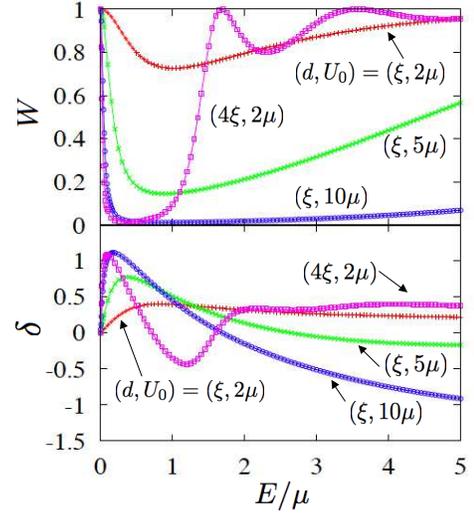}}
\caption{(Color online). Calculated transmission probability $W$ (upper panel) and phase
 shift $\delta$ (lower panel). We take the rectangular
 potential barrier with $(d,U_0)=(\xi,2\mu)$, $(\xi,5\mu)$, $(\xi,10\mu)$,
 and $(4\xi,2\mu)$.}
\label{fig1}
\end{figure}

Figure~\ref{fig1} shows the calculated transmission probability
$W\equiv|t|^2$, as well as the phase shift $\delta\equiv {\rm arg}(t)$
as functions of the incident energy $E$. 
The anomalous tunneling behavior discussed in \cite{Kovrizhin,Kagan} can
be clearly seen in Fig.~\ref{fig1}, i.e., $W\to1$ and $\delta\to 0$ when $E\to0$. 
Around $E=0$, one can see the enhancement of $W$, the region of which is
wider for weaker potential barrier. 
When the incoming energy $E$ is very large ($E\gg\mu$), since the
Bogoliubov phonon loses its collective nature, the tunneling property is
close to that of a single particle.

The upper panel in Fig. \ref{fig1} shows the resonance
tunneling behavior ($W=1$) at finite energies for
$(d,U_0)=(4\xi,2\mu)$. 
At each resonance energy, $u$ has a large amplitude in the
barrier while $v$ is suppressed. This is quite different from the case
of anomalous tunneling at $E\simeq 0$, 
where $u$ and $v$ monotonically decrease under the barrier
approaching the zero-mode $\Psi_0$.
It was proposed that the anomalous tunneling is due to the
quasiresonance scattering by the potential wells formed near the
barrier \cite{Kagan}. However, we do not find any signature of it in $u$
and $v$.

To see the low energy quasiparticle transmission in more detail, we
directly calculate the quasiparticle current $J_{\rm q}$. First, we
discuss current conservation for Bogoliubov phonons. In the Bogoliubov
theory, the total number density
$n=\langle\hat\psi^\dagger\hat\psi\rangle$ and current $J=(1/m){\rm
Im}\langle\hat\psi^\dagger\partial_x\hat\psi\rangle$ are respectively given by
\begin{eqnarray}
n=n_{\rm s}+\sum_j\left(n_{u_j}+n_{v_j}\right)\langle\hat\alpha_j^\dagger\hat\alpha_j\rangle+\sum_jn_{v_j},\label{density}
\\
J=J_{\rm s}+\sum_j\left(J_{u_j} - J_{v_j}\right)
\langle\hat\alpha_j^\dagger\hat\alpha_j\rangle-\sum_j J_{v_j}.
\label{current}
\end{eqnarray}
Here, $n_{u_j}=|u_j|^2$, $n_{v_j}=|v_j|^2$, 
$J_{u_j}=(1/m){\rm Im}(u_j^\ast\partial_x u_j)$, 
and $J_{v_j}=(1/m){\rm Im}(v_j^\ast\partial_x v_j)$, 
where $n_{\rm s}=|\Psi_0|^2$ is the condensate density, 
and $J_{\rm s}=(1/m){\rm Im}(\Psi_0^\ast\partial_x \Psi_0)$ is the
supercurrent density associated with the phase-twisted condensate
wavefunction. The total number density $n$ and total current $J$ satisfy
the continuity equation $\partial_t n+\partial_xJ=0$. 
Since the second terms in Eqs.~(\ref{density}) and (\ref{current})
describe the quasiparticle contributions, the quasiparticle density
and quasiparticle current are, respectively, given by $n_{\rm
q}=n_u+n_v$ and $J_{\rm q}=J_u-J_v$. 
In a uniform system, the creation of a Bogoliubov phonon with momentum $k$
induces the quasiparticle current $J_{\rm q}=k/(mV)$. 
(Note that $J_u=(k/m)a^2$, $J_v=(k/m)b^2$.)
The last term in Eq.~(\ref{density}) is the so-called quantum depletion
at $T=0$, which is the non-condensate density originating from the
repulsive interaction between bosons.

Using the time-dependent Bogoliubov equations,
\begin{eqnarray}
i\tau_3\partial_t
\left(\begin{array}{l}
u\\
v
\end{array}\right)=
\left(
\begin{matrix}
\hat{h} & -g\Psi_0^2\\
-g(\Psi_0^\ast)^2 & \hat{h}
\end{matrix}
\right)
\left(
\begin{array}{l}
u\\
v
\end{array}
\right),
\label{TDBogoliubov}
\end{eqnarray}
one obtains the continuity equations for $u$ and $v$, as
$\partial_tn_u+\partial_xJ_u=S/2$ and $\partial_t n_v-\partial_x J_v=S/2$,
where the source term has the form $S=-4g{\rm Im}\left(\Psi_0^2u^\ast v
\right)$. Thus, the continuity equation for quasiparticles
is obtained as
\begin{eqnarray}
\partial_t n_{\rm q}+\partial_x J_{\rm q}=S\ .
\label{qcontinuity}
\end{eqnarray}
In a uniform case, one finds $S=0$, so that the number of
quasiparticles is conserved. On the other hand, as will be shown later,
since the source term $S$ is finite near the barrier in our
inhomogeneous problem, the number of quasiparticles is {\it not}
conserved. In particular, in the stationary state ($\partial_t n_q=0$),
Eq.~(\ref{qcontinuity}) indicates that there is a source supplying excess
quasiparticle current in addition to the incoming current injected from
$x=-\infty$. 

In contrast to the non-conserved quasiparticles, the continuity
equation for energy density $\rho_{\rm q}\equiv E(n_u-n_v)$ has no
source term, as 
$\partial_t\rho_{\rm q}+\partial_x Q_{\rm q}=0$ \cite{Kagan}, where 
$Q_{\rm q}=E(J_u+J_v)$ is the energy flux carried by quasiparticles. 
Namely, $Q_{\rm q}$ is conserved in the stationary state. 
Using the asymptotic form in Eq.~(\ref{asympt}), we obtain $Q_{\rm
q}=E(k/m)(a^2+b^2)(1-|r|^2)$ and $J_{\rm q}=(k/mV)(1-|r|^2)$  for
$x \to -\infty$,  and $Q_{\rm q}=E(k/m)(a^2+b^2)|t|^2$ and $J_{\rm
q}=(k/mV)|t|^2$ for $x\to \infty$. 
In the stationary state, noting that $Q_{\rm q}$ is conserved, one
obtains $|r|^2+|t|^2=1$. From this result, we find that
$J_{\rm q}(x=-\infty)=J_{\rm q}(x=\infty)$.

\begin{figure}
\centerline{\includegraphics[width=6cm]{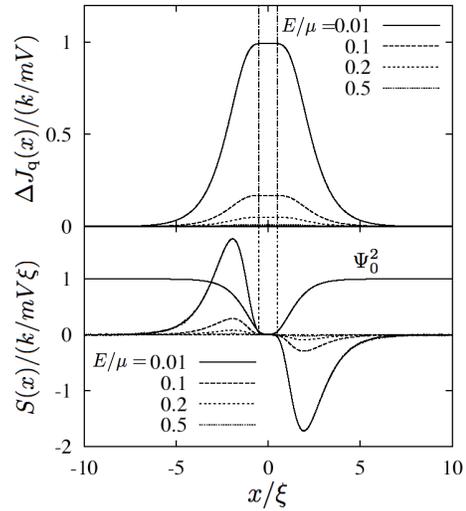}}
\caption{Upper panel: Quasiparticle current $\Delta J_{\rm q}(x)\equiv
 J_{\rm q}(x)-J_{\rm q}(-\infty)$ measured from the value at
 $x=-\infty$. Lower panel: Spatial variation of the source term $S$. We
 take the rectangular barrier with $(d,U_0)=(\xi,10\mu)$. 
The dash-dotted line indicates the potential barrier $U(x)$.}
\label{fig2}
\end{figure}

The upper panel in Fig.~\ref{fig2} shows the {\it excess} quasiparticle
current $\Delta J_{\rm q}(x)\equiv J_{\rm q}(x)-J_{\rm q}(\infty)$. 
Near the barrier, the quasiparticle current is enhanced, which is
remarkable in the low energy region. This enhancement occurs when the
source term $S$ is finite, as shown in the lower panel in
Fig. \ref{fig2}. Indeed, from Eq.~(\ref{qcontinuity}), we find that
$\Delta J_{\rm q}$ is related to $S$ as 
$\Delta J_{\rm q}(x)=\int_{-\infty}^x S(y) dy$.  
The source term literally works as a source when $S>0$, while it works
as a drain when $S<0$. In the barrier region ($|x|\le d/2$), the source
term vanishes, so that the magnitude of the enhanced quasiparticle
current is constant, as shown in the upper panel in Fig. \ref{fig2}. 
We note that the excess component $\Delta J_{\rm q}$ of the quasiparticle
current is completely absorbed by the source term in the right hand side
of the barrier. 
Thus, the magnitude of the outgoing quasiparticle current far away from
the barrier reduces to that of incident quasiparticle current, as
expected. In the low energy limit, the maximum excess quasiparticle
current $\Delta J_{\rm q}(x=0)$ is proportional to $k$. Thus, $\Delta
J_{\rm q}(0)/(k/mV)$ in Fig.~\ref{fig2} approaches a constant height which
becomes larger for larger potential barrier.

The transmission probability of Bogoliubov phonon
increases when the quasiparticle current is enhanced near the
barrier. Comparing Fig.~\ref{fig2} with the result for
$(d,U_0)=(\xi,10\mu)$ in Fig.~\ref{fig1}, one finds that the energy
($E\sim 0.1\mu$) where the high transmission probability associated with
the anomalous tunneling becomes remarkable coincides with the energy
where the excess current in the barrier (normalized by $k/mV$) becomes large. 

The excess quasiparticle current is supplied by the condensate. To
see this, we note that the divergence of Eq.~(\ref{current}) in the
stationary state gives $\partial_x J=\partial_x J_{\rm s}+\sum_j
S_j\langle\hat\alpha_j^\dagger\hat\alpha_j\rangle+\frac{1}{2}\sum_j S_j$.
Since the supercurrent is conserved ($\partial_xJ_{\rm s}=0$), 
it reduces to $\partial_x
J=\frac{1}{2}\sum_jS_j$ in the ground state
($\langle \hat\alpha_j^\dagger\hat\alpha_j\rangle=0$). 
Since the total current $J$ is conserved, the sum of the source term vanishes ($\sum_j S_j=0$). 
When quasiparticles are excited
($\langle\hat\alpha_j^\dagger\hat\alpha_j\rangle\neq 0$),
one obtains $\partial_x J\ne 0$, which contradicts
with the conservation of the total current. 
This inconsistency is actually eliminated when one includes effects of 
quasiparticles on the condensate, as 
\begin{eqnarray}
\left(-\frac{1}{2m}\frac{d^2}{dx^2}+U(x)+g|\Psi_0|^2\right)\Psi_0&-&2g\sum_j
 u_jv_j^\ast\langle\hat\alpha_j^\dagger\hat\alpha_j\rangle\Psi_0^\ast\nonumber\\
&=&\mu\Psi_0\ .
\label{gGP}
\end{eqnarray}
In this GP equation, the last term on the left side is the
quasiparticle contribution, which originates from the
so-called anomalous average term
$g\langle\delta\hat\psi\delta\hat\psi\rangle\Psi_0^\ast$ (which is
neglected in deriving the GP equation within the Bogoliubov
approximation \cite{Griffin,Stoofcomment}). Using Eq.~(\ref{gGP}), we find that
the continuity equation for $J_{\rm s}$ is modified as $\partial_x J_{\rm s}
=-\sum_jS_j\langle\hat\alpha_j^\dagger\hat\alpha_j\rangle$.
This result gives
the expected conservation of the total current $\partial_x J=0$ in the
presence of Bogoliubov phonons. This justifies the
modification of the GP equation shown in Eq.~(\ref{gGP}).

Using $S_j=\partial_x J_{{\rm q}j}$ and integrating the new continuity
equation for $J_{\rm s}$ in
terms of $x$ from $-\infty$ to $x$, we obtain $\Delta J_{\rm
s}=-\sum_j\Delta J_{{\rm
q}j}\langle\hat\alpha_j^\dagger\hat\alpha_j\rangle$, 
where $\Delta J_{\rm s}(x)$ is the deviation of supercurrent from the
value based on Eq.~(\ref{GP}). This equation shows that the
sum of supercurrent and quasiparticle current is always
conserved. Thus, the excess quasiparticle current shown in the upper
panel in Fig. \ref{fig2} is found to be supplied from the condensate. At
the same time, the quasiparticle current also affects the supercurrent.
Since the quasiparticle current is enhanced near the barrier, the induced
supercurrent flows only near the barrier, in the opposite direction to
the quasiparticle current to conserve the total current.

\begin{figure}
\centerline{\includegraphics[width=6cm]{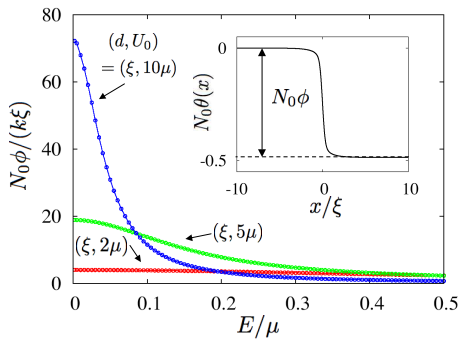}}
\caption{(Color online). Phase difference $\phi$ across the barrier as a function of the
 incident energy $E$. $N_0=n_0V$ is the number of condensate atoms. We
 take $U_0=2\mu$, $5\mu$, and
 $10\mu$ with a fixed barrier width $d=\xi$. 
 The inset shows the phase $\theta(x)$ for $(d,U_0)=(\xi,10\mu)$ and
 $E=0.01\mu$.}
\label{fig3}
\end{figure}

If we assume that the influence of quasiparticle changes the condensate
wavefunction by a phase shift $\theta(x)$, the induced supercurrent is
given by $\Delta J_{\rm s}=(1/m)n_{\rm s}(\partial_x\theta)$ \cite{phase}. 
Thus, we find that $\theta(x)$ is given by
\begin{equation}
\theta(x)=-m\sum_j\left(\int_{-\infty}^x dy \frac{\Delta J_{{\rm
		   q}j}(y)}{n_{\rm
		   s}(y)}\right)\langle\hat\alpha_j^\dagger\hat\alpha_j\rangle\ .
\end{equation}
The inset in Fig.~\ref{fig3} shows the spatial variation of $N_0\theta(x)$, where $N_0\equiv n_0V$
is the number of condensate atoms. 
Here, we assume
that only one Bogoliubov phonon with energy $E$ exists. 
The phase $\theta(x)$ sharply varies around the barrier.
This clearly shows that the Bogoliubov phonon twists the relative phase
of the condensates on the left and right sides of the barrier.
Since their coupling is weak at the barrier region, the relative phase
is subject to change by the tunneling quasiparticle.
The induced supercurrent $\Delta J_{\rm s}$ can be naturally regarded
as the Josephson current originating from the phase difference
$\phi\equiv\theta(x=-\infty)-\theta(x=\infty)$.

Figure~\ref{fig3} shows the normalized phase difference $N_0\phi/(k\xi)$
as a function of $E$. It is enhanced at low
energies similarly to the normalized excess quasiparticle current
$\Delta J_{\rm q}/(k/mV)$ in Fig.~\ref{fig2} and indeed, normalized
supercurrent $\Delta J_{\rm s}/(k/mV)$. 
Actually, the maximum supercurrent $\Delta J_{\rm s}(x=0)$ is
proportional to $\phi$ at low energies. 
This result justifies regarding $J_{\rm s}$ as the Josephson current due to $\phi$.
In addition, Fig.~\ref{fig3} shows that, at the same
energy, $\phi$ is larger for higher barriers. 
This implies that, since the coupling of the condensates across the barrier
becomes weaker as the barrier gets higher, the phase of the condensate wavefunction
is easily twisted by the Bogoliubov phonon for higher barriers.

In summary, we have investigated tunneling effect of Bogoliubov
phonon in a BEC at $T=0$. 
We found anomalous tunneling properties of Bogoliubov phonons.
The quasiparticle current of Bogoliubov phonon is enhanced near
the potential barrier due to the supply from the condensate.
This enhancement is remarkable at low incident energies, and explains
the increase of the transmission probability of Bogoliubov phonon in \cite{Kovrizhin, Kagan}.
Furthermore, we have shown that the quasiparticle current twists the phase of the
condensate wavefunction across the barrier, which induces the
counterflow of Josephson supercurrent to conserve the total current.
This twisted phase could be observed in the interference pattern when
a BEC is released from the trap potential as in \cite{Andrews2}.

S.T. acknowledges I. Danshita, K. Kamide, S. Inouye, and F. Dalfovo for useful discussions.


\begin{thebibliography}{}
\bibitem{Stoofcomment}The anomalous average term has a quantum
	contribution $-g\sum_ju_jv_j^\ast\Psi_0^\ast$, which yields the
	additional term $\frac{1}{2}\sum_jS_j$ in the continuity
	equation. In this paper, we treat the problem perturbatively in
	the sense that $\Psi_0$ and $(u_j,v_j)$ 
	obtained from Eq.~(\ref{GP}) and (\ref{Bogoliubov}) are used in
	evaluating backreaction effect of quasiparticles on condensate,
	i.e. the correction to $\Psi_0$.
	Within this treatment, one finds $\sum_jS_j=0$. As a result, the
	quantum contribution in the continuity 
	equation vanishes. To improve this, one needs
	to solve the generalized GP and Bogoliubov equations including
	the anomalous average terms self-consistently
	(Hartree-Fock-Bogoliubov approximation). However, it yields an
	excitation spectrum with a finite energy gap which is unphysical
	as discussed in \cite{Griffin}.
\bibitem{phase}This assumption is valid as long as $\theta(x)$ is
	small. Since $\theta(x)$ is inversely proportional to the number
	of condensate atoms $N_0$, it is negligibly small. In this case, 
        $\Psi_0$ can be obtained without solving Eq.~(\ref{gGP}).
\end{thebibliography}
\end{document}